\documentclass[twocolumn,showpacs,preprintnumbers,aNPSath,amssymb,pre,nofootinbib]{revtex4-1}

\usepackage[utf8]{inputenc}
\usepackage[T1]{fontenc}
\usepackage{multirow}
\usepackage{mathptmx}
\usepackage[export]{adjustbox}
\usepackage{mathtools}
\usepackage{float}
\usepackage{color}    
\usepackage{graphicx}
\usepackage{dcolumn}
\usepackage{bm}
\usepackage{subfigure}
\usepackage{amssymb}
\usepackage{amsmath}

\usepackage{siunitx}
\DeclarePairedDelimiterXPP\BigOSI[2]%
  {\mathcal{O}}{(}{)}{}%
  {\SI{#1}{#2}}
  
\begin{document}

\title{Impact of the local field correction on transport and dynamic properties of warm dense matter}

\author{S.K. Kodanova}
\affiliation{ Institute for Experimental and Theoretical Physics, Al-Farabi Kazakh National University, 71 Al-Farabi ave., 050040 Almaty, Kazakhstan}

\author{T.S. Ramazanov}
\affiliation{ Institute for Experimental and Theoretical Physics, Al-Farabi Kazakh National University, 71 Al-Farabi ave., 050040 Almaty, Kazakhstan}

\author{M.K. Issanova}
\email{issanova@physics.kz}
\affiliation{ Institute for Experimental and Theoretical Physics, Al-Farabi Kazakh National University, 71 Al-Farabi ave., 050040 Almaty, Kazakhstan}


\begin{abstract}
The plasma screening model, taking into account the electronic exchange-correlation effects and the ionic non-ideality in dense quantum plasmas, is presented. This model can be used as an input in various plasma interaction models to calculate the scattering cross-sections and transport properties. The application of the presented plasma screening model is demonstrated on the example of the temperature relaxation rate in dense hydrogen and warm dense aluminum. Additionally, we compute the conductivity of warm-dense aluminum in the regime where collisions are dominated by electron-ion scattering. The obtained results are compared with available theoretical results and simulation data.

\end{abstract}

\maketitle

\section{\label{Sec. I} Introduction}

Investigating matter at extremely high temperatures and densities has attracted considerable interest from researchers in recent years. 
Experimental study of warm dense matter (WDM) at such laser facilities as NIF \cite{Tilo_Nature_2023}, European XFEL \cite{Zastrau_2021}, and OMEGA \cite{Falk_HEDP_2012}
 opens a path to reveal the internal structure of planets down to their cores, both in our solar system and beyond \cite{Saumon}. Also, these high-pressure experiments will allow us to test theories of states of matter under extreme conditions of density and temperature \cite{Guillot, Nellis, Burrows, Tahir2, Moldabekov_Omega_2024, Moldabekov_PRR_2024}. 
This attention is prompted not only by the significance of the physical properties of dense plasma for advancing fields like astrophysics \cite{Baraffe}, inertial confinement fusion \cite{Saumon}, and laser technology \cite{Moldabekov_Nat, Moses, Silva}, but also by the potential practical applications such as the synthesis of novel materials \cite{Kraus2016, Kraus2017}.

Under experimental conditions, plasma and heated dense matter are created by rapid energy release processes and subsequent relaxation. Because, as a rule, plasmas are created with the initial state being out of equilibrium, understanding temperature relaxation is critical for modeling the evolution of multi-temperature high energy-density plasmas \cite{Cho, Simoni, Simoni2, Vorberger2010}.  In this regard, directly and accurately measuring relaxation rates is challenging due to ultrashort time scales and complex initial conditions.
Therefore, a theoretical description of the relaxation and relevant transport properties is necessary to interpret these experiments. Temperature relaxation has been theoretically studied extensively for electron-ion systems \cite{Vorberger1, Dharma-wardana, Vorberger2, Gericke}. These theories have been compared to MD simulations with varying degrees of success \cite{Glosli, Dimonte}. However, explicit electron-ion MD simulations often rely on quantum statistical potentials \cite{Jones, Murillo8}, which may only be valid in thermodynamic equilibrium. This complicates comparisons of MD simulations with theory because disagreements can be attributed to uncertainties in the interaction potentials instead of theoretical models. Carefully designed and accurately diagnosed laboratory experiments are required to test the reliability of these theoretical approaches. One such experiment is presented in Ref. \cite{Sprenkle}, where ion-ion temperature relaxation in a binary mixture was studied by exploiting dual-species ultracold neutral plasma and compared measured relaxation rates with atomistic simulations and a range of popular theories. 

First-principles computational modeling methods, such as density functional theory (DFT) \cite{Burke, Galli, Collins} and the quantum Monte Carlo method (QMC) \cite{Bowen, Moroni, Ceperley}, are indispensable for the reliable calculation of WDM properties. For example, the QMC results are used as an input for various approximations \cite{Perdew}. For WDM applications, Groth \textit{et al.,} \cite{Groth2} provided accurate ab initio QMC results for exchange-correlation free energy for various densities, temperatures, and spin polarization degrees.

In addition to ab into methods, various approximate models are still relevant for describing the transport properties of WDM. This is because of the high computational costs of ab initio methods, such as the density functional theory at high temperatures. For example, the Boltzmann equation with the Landau or Lennard-Balescu collision integral can be used to calculate the transport coefficients. One of the methods for solving the kinetic equation is the Chapman-Enskog theory \cite{Chapman}, based on expansion in orthogonal Sonin polynomials. For fully ionized plasma, including electron-electron collisions in the Chapman-Enskog method, Spitzer and H\"{a}rm \cite{Spitzer} were first to provide an adequate analytical result for the plasma conductivity. Often, plasma properties are evaluated using the Debye-Hückel potential for classical plasma and the Thomas-Fermi potential for degenerate ideal electrons \cite{Moldabekov2, Moldabekov3, Moldabekov4, Stanton}. There are several modifications of these potentials taking into account various quantum and non-ideality effects \cite{POP2021, POP2022, CPP2024, Gabdullin, Kodanova1, Kodanova2, Kodanova3, Moldabekov1, Issanova, MRE18, Moldabekov19}.  These potentials have been widely used to study various characteristics of nonideal plasmas \cite{Gregori, Das_pre_2020, Chen, Ramazanov, Arkhipov2, Arkhipov3} and better agreement with experimental data has been obtained in the temperature and density ranges studied. Many works have studied electron-ion scattering in the dense plasma regime, taking into account charge screening in partially and weakly degenerate cases \cite{Becker, Daligault, Hu, Brovman}.

\begin{figure*}[ht]
    \centering
\includegraphics[width=1.0 \textwidth]{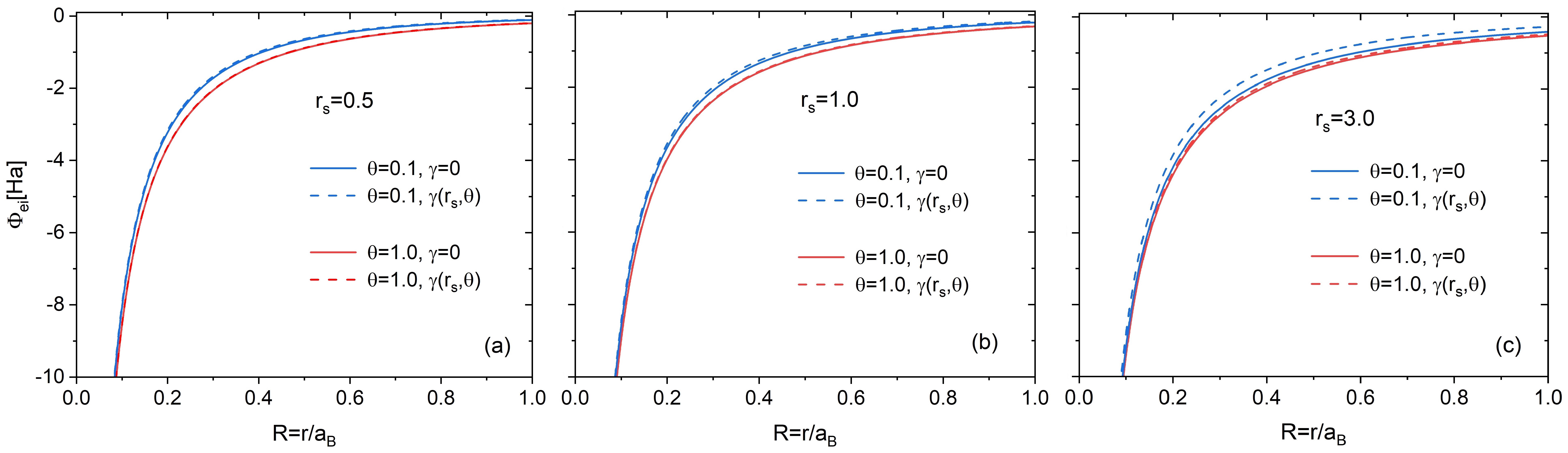}
\caption{Effective potential of interaction of an electron with a hydrogen ion with and without the electronic non-ideality correction ($\gamma$) at a fixed density parameters (a) $r_{s}=0.5$, (b) $r_{s}=1$, (c) $r_{s}=3$ for different values of the degeneracy parameter $\theta=0.1$ and $\theta=1$.}
\label{Fei}
\end{figure*}

A key approach for computing the transport properties of many-particle systems is linear response theory. Linear response theory can be used to analyze how a plasma responds to small changes in external conditions or fields. For example,  in Ref. \cite{Holst}, using Kubo linear response theory, the authors computed electrical conductivity and thermal conductivity in a strongly correlated electronic system within the framework of ab initio molecular dynamics simulation. In Ref. \cite{Arkhipov3},  an approximate expression for the generalized Coulomb logarithm based on the Ornstein-Zernike integral equation in the hyperchain approximation with different interaction potentials was  obtained. They considered electron-ion plasma as a classical system of pseudoparticles interacting through those effective pair potentials that mimic the effects of quantum diffraction at short distances.
In works \cite{Hu, French, Desjarlais},  an average atom model based on DFT was used to compute the potential of average force, which is determined by solving the Ornstein–Zernike equations and takes into account the correlations between ions and valence electrons \cite{Starrett}. This approach provided a good agreement with experimental electrical conductivity.
In Refs. \cite{Reinholz, Röpke2, Röpke3}, the expressions were obtained for the electrical conductivity and absorption coefficients of high-temperature plasma based on linear response theory. We note that within the framework of the virial expansion of the electrical conductivity of a fully ionized plasma, which takes into account many-particle effects, various limiting cases are considered, and the corresponding interpolation formulas for the electrical conductivity coefficient are given \cite{Röpke} and compared with experimental values.

\begin{figure}[ht]
\includegraphics[width=0.85 \linewidth]{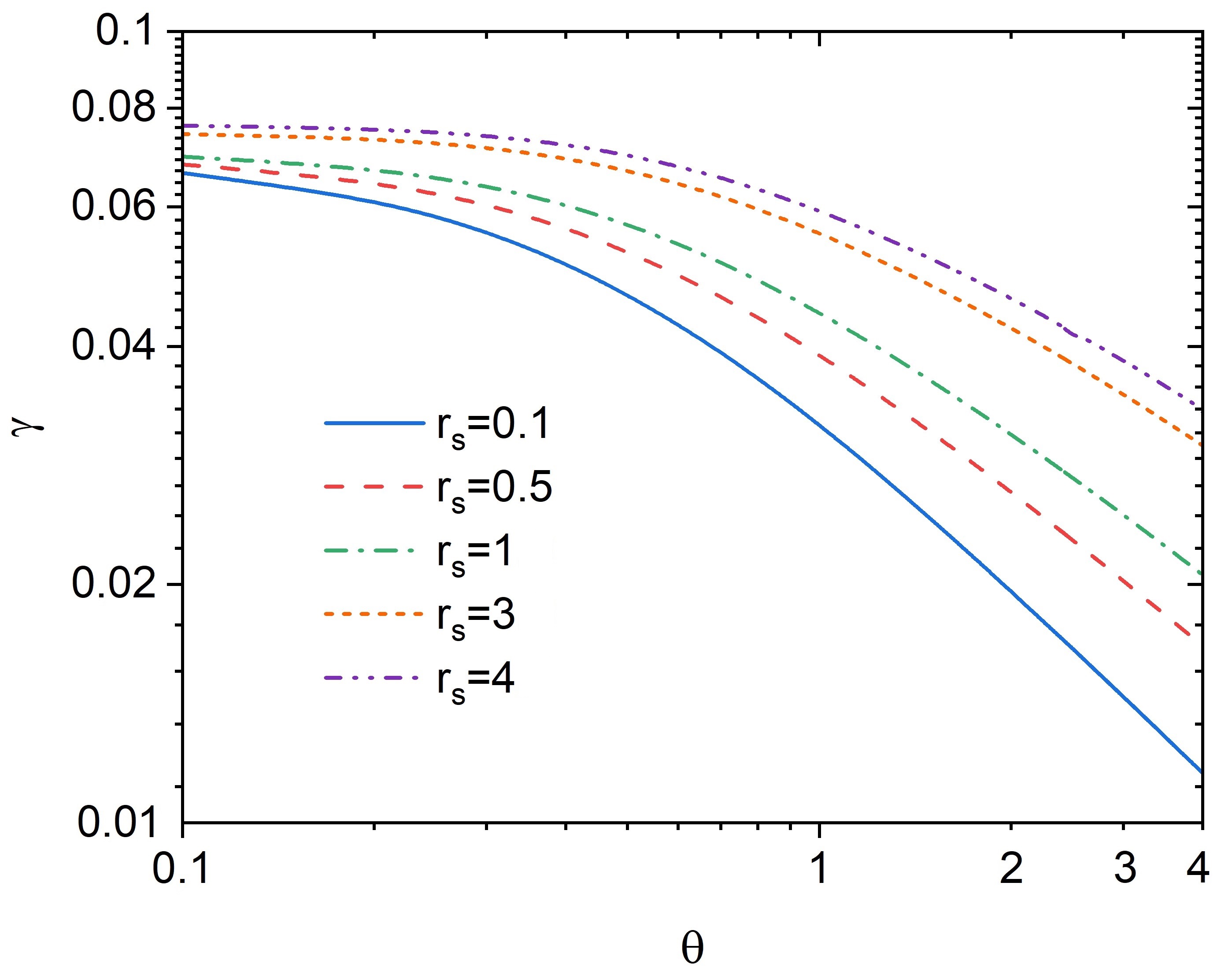}
\caption{Correction to electronic screening from Eq.~(\ref{LFCgamma}) depending on the degeneracy parameter for various parameters $r_{s}$.}
\label{gamma}
\end{figure}

For an adequate description of WDM and dense plasmas, it is critical to take into account the exchange-correlation effects.
In an earlier series of studies by Ichimaru, Mitake, Tanaka, and Yan \cite{Ichimaru, Ichimaru2, Ichimaru3} it was shown that in a dense plasma, exchange-correlation effects between charged particles become essential. To go beyond the random phase approximation (RPA), the local field correction within linear response formalism is used.
Utilizing local field correction, various analytical formulas for screened potentials using the long-wave approximation approach and taking into account exchange-correlation effects were analyzed in Refs. \cite{Moldabekov18, Moldabekov_APL, Bonitz20, Dornheim21}.

In this work, we use the analytical parameterization of the QMC results for the free energy density of uniform electron gas in WDM regime \cite{Groth2} in combination with a simple screened Thomas-Fermi-type potential to compute the electron-ion scattering cross sections and generalized Coulomb logarithm of dense hydrogen plasmas. In addition to electronic contribution, the presented work-flow takes into account the screening due to ions using a simple receipt suggested by Murillo \textit{et al.,} \cite{PhysRevE.93.043203}.
We show that our simple and fast computational scheme allows us to achieve remarkably good agreement with the first-principles DFT results.

\section{\label{Sec. II} Theory}
\subsection{Plasma screening model}

In this work, to study the dynamic properties of WDM plasma, we used the Yukawa-type potential and the classical Chapman-Enskog method to calculate the Coulomb logarithm based on the transport scattering cross-section. Particle scattering cross-sections were determined from the Calogero equation for scattering phases.

The main goal of the presented work is to quantify the influence of electron degeneracy and the effect of electron exchange correlations on the temperature relaxation process and the transport properties of WDM plasma compared to what is used in standard plasma physics. A similar analysis can be carried out for other properties, such as kinetic and optical properties.

\begin{figure}[ht]
\includegraphics[width=0.9\linewidth]{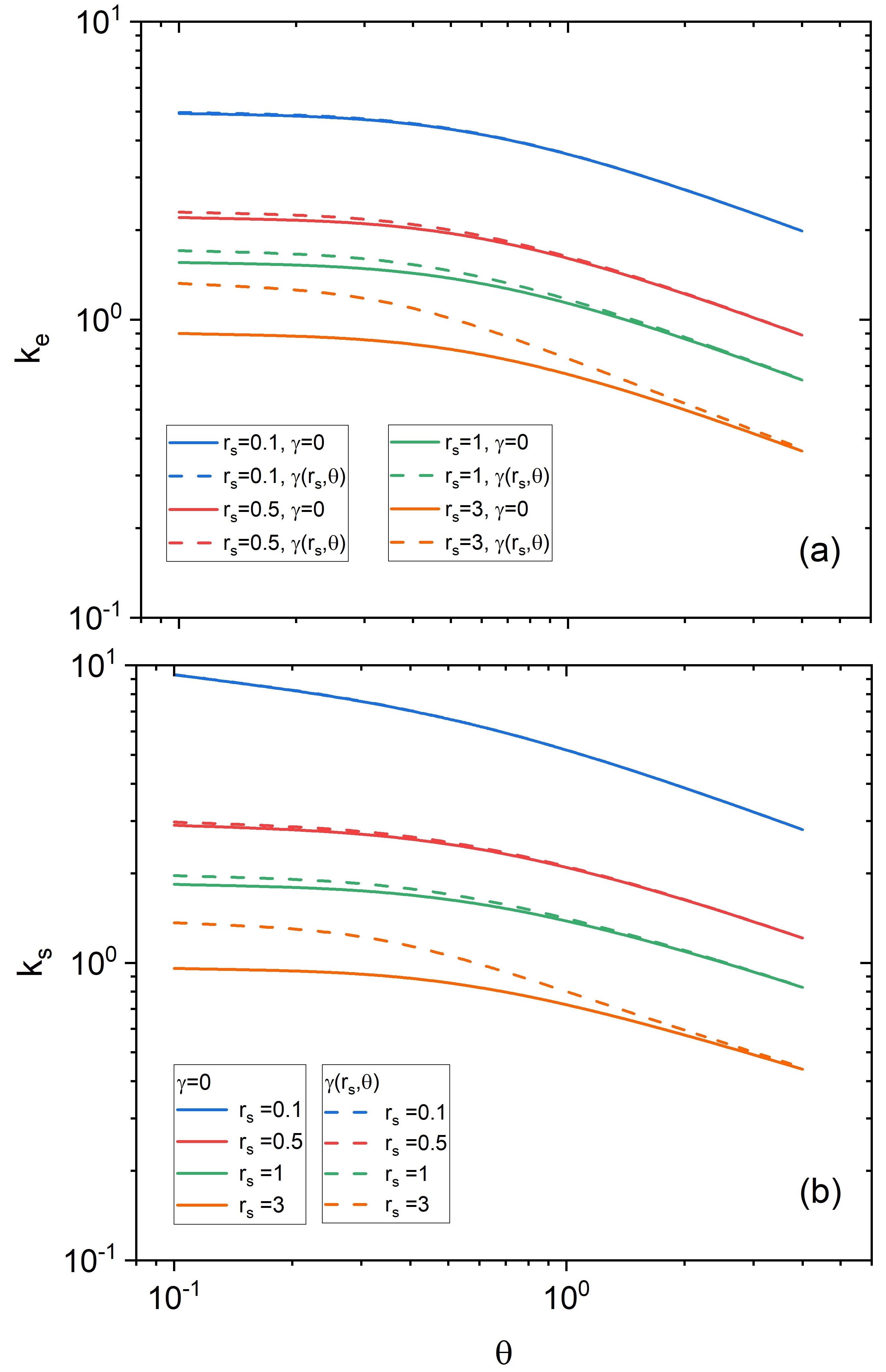}
\caption{Inverse screening length $k_{e}$ in units $a_{B}$ depending on the degeneracy parameter $\theta$ with and without taking into account exchange-correlation effects for various parameters $r_{s}$.}
\label{ksfigs}
\end{figure}

The used screened electron-ion potential reads
\begin{equation}\label{Yukawa}
\Phi_{Y} =\frac{Ze}{r} \exp(-k_{S} r ),
\end{equation}
with the inverse screening parameter it the form \cite{MRE18}:
\begin{equation}\label{eq:ks_tot}
    k_{s}^{2}=k_{e}^{2}(T_e,n_e)+k_{i}^{2}(T_i,n_i),
\end{equation}
where $k_{e}$ and $k_{i}$ are the contribution of electrons and ions to the screening, accordingly. We consider a general case where the temperature of electrons $T_e$ and ions $T_i$ are not equal. Besides of temperatures, the screening parameter depends on the density of electrons $n_e$ and ions $n_i$.

Potential (\ref{Yukawa}) follows from using the long wavelength limit of 
the density response function of the uniform electron gas \cite{Moldabekov2, MRE18, Moldabekov3, Moldabekov18}. 

The quality of the calculation results using potential (\ref{Yukawa}) depends on the plasma screening model used to calculate the screening parameter $k_s$.  Moldabekov \textit{et al} \cite{Moldabekov2} have shown that the electron screening parameter of the UEG gas has the following form: 
\begin{equation}\label{KS}
k_{e}^{-2}(T_e,n_e) =k_{id}^{-2}(T_e,n_e)-k_{F}^{-2}(n_e)\gamma(T_e,n_e),
\end{equation}
where $k_{F}(n_0)$ is the Fermi wavenumber, and $k_{id}(T_e,n_e)$ denotes the screening parameter of the ideal UEG:
\begin{equation}
    k_{id}^{2}(T_e,n_e) =k_{TF}^{2}(n_e) \theta ^{{1\mathord{\left/ {\vphantom {1 2}} \right. \kern-\nulldelimiterspace} 2} } I_{{-1\mathord{\left/ {\vphantom {-1 2}} \right. \kern-\nulldelimiterspace} 2} } (\eta )/2,
\end{equation} 
and $k_{TF}^{2}(n_e)=(\sqrt{3}\omega_{p})/v_F$ ist the Thomas-Fermi screening parameter, $\theta=k_BT_e/E_F$ is the degeneracy parameter of elctrons, $I_{{-1\mathord{\left/ {\vphantom {-1 2}}\right.\kern-\nulldelimiterspace} 2} } (\eta )$ is the Fermi ontegral of order $-1/2$ at the chemical potential of electrons $\eta=\mu/(k_BT)$ (see Ref. \cite{PhysRevA.29.1471}).

\begin{figure*}[ht]
    \centering
\includegraphics[width=1.0 \linewidth]{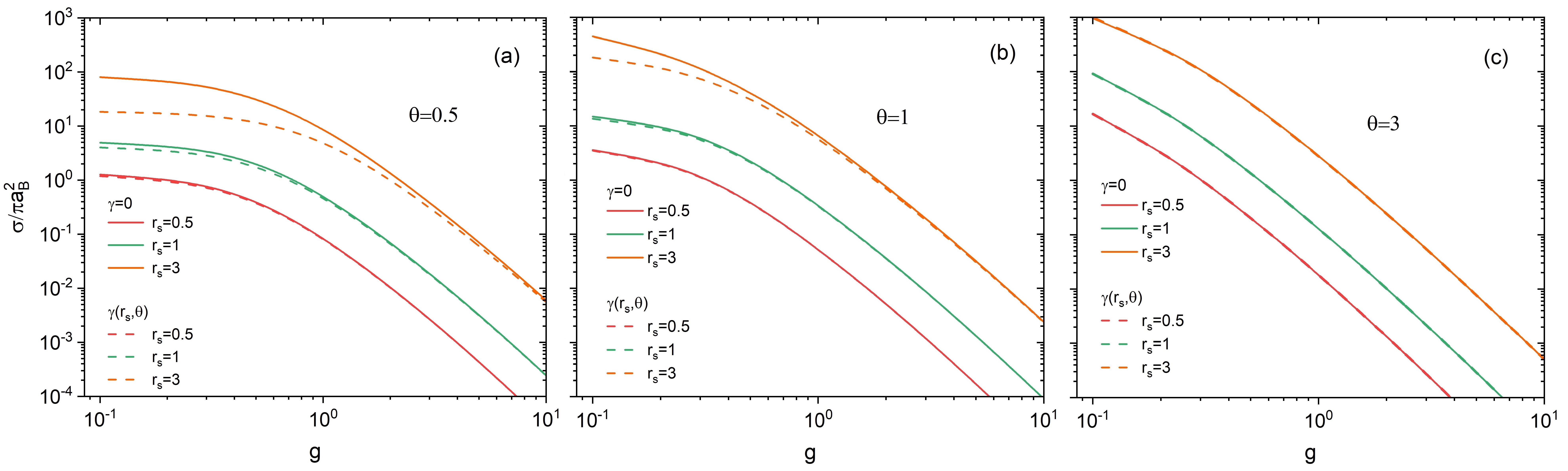}
\caption{Transport cross sections of electron-ion scattering with a fixed degeneracy parameter (a) $\theta=0.5$, (b) $\theta=1$, (c) $\theta=3$ at different values of the density parameter $r_{s}$ for cases with and without the electronic non-ideality correction $\gamma$.}
\label{crosssections}
\end{figure*}

The dimensionless coefficient $\gamma(T_e,n_e)$ in  Eq.~(\ref{KS}) takes into account the correction to the electronic screening due to electronic exchange-correlations (non-ideality) \cite{Moldabekov2, Moldabekov19, Amirov}
and represents the contribution to the electronic density response function due to the static local field correction in the long-wavelength limit \cite{Moldabekov2022-gb}:
\begin{equation}\label{EqG}
 G(k)\simeq \gamma (k^2/k_F^2)+....,   
\end{equation}
which also defines the exchange-correlation kernel of electrons \cite{Moldabekov2023-lt}.

The value of $\gamma(T_e,n_e)$ is computed using compressibilty sum-rule \cite{Moldabekov2}:
\begin{equation}\label{LFCgamma}
    \gamma(T_e,n_e)= -\frac{k_F^2}{4\pi e^2}\frac{\partial^2 [n f_{\rm xc}(T_e,n_e)]}{\partial n^2},
\end{equation}
where $k_F=v_F m/\hbar$, and  $f_{\rm xc}$ is exchange-correlation free energy per electron of the UEG from the Quantum Monte-Carlo simulations based parameterization presented in Ref.~\cite{Groth2}.

Despite having a simple form, Moldabekov \textit{et al} demonstrated that the electron screening model (\ref{KS}) provides an accurate description of the test charge screening in plasmas compared to the model using the exact static density response function of the UEG \cite{Moldabekov2}. This was also demonstrated by Moldabekov \textit{et al} \cite{Moldabekov_APL, Moldabekov_pre_2019} by performing molecular dynamic simulations of the static and dynamic structure factors of ions using various screening models.

The next ingredient needed to complete the plasma screening model is the contribution of ions $k_{i}(T_i,n_i)$ in Eq. (\ref{eq:ks_tot}).
Stanton and Murillo \cite{Stanton} performed a detailed analysis of the screening due to ions at different ionic non-ideality regimes. 
They showed that the following simple model provides an accurate description in a wide range of parameters \cite{Stanton}:
\begin{equation}\label{ki}
k_{i}^{2}=1/(\lambda_{i}^{2}+a_{i}^{2}),
\end{equation}
where $\lambda_{i}=(k_{B}T_{i}/4\pi n_{i}e^2 Z^2)^{1/2}$ is the ideal ion screening length and $a=(3/4\pi n)^{1/3}$ is the ion sphere radius (mean distance between ions). In the case of multicomponent plasmas, the charge density of ions is computed as $n_i=\sum_{\alpha}n_{\alpha}$, with $\alpha$ denoting the type of ions in plasmas. We note that in our notation, the charge density takes into account the charge state of ions. This is in contrast to the number density of ions. 

Eqs. (\ref{eq:ks_tot})-(\ref{ki}) define our plasma screening model for dense plasmas across plasma degeneracy and coupling regimes.  
Being based on the results of the analyses performed by Moldabekov \textit{et al} \cite{Moldabekov2, Moldabekov3, Moldabekov_APL, Moldabekov_pre_2019} and by Stanton and Murillo \cite{Stanton}, we refer to this model as the nonideal plasma screening (NPS) model. 
We use the NPS model to compute the generalized Coulomb logarithm in dense plasmas \cite{Rightley}, which allows us to calculate the temperature relaxation rate and conductivity.

\subsection{Generalized Coulomb logarithm}

Following a recent work by Rightley and Baalrud \cite{Rightley},
the generalized Coulomb logarithm is computed using the transport cross-section $Q^{T}$ defined by electron-ion collisions:
\begin{equation} \label{Xi}
 \Xi= {\frac{1}{2} \int_{0}^{\infty} dg G(g) \frac{Q^{T}(g)}{\sigma_{0}}},
\end{equation}
where $\sigma_{0} =\hbar ^{2} /(m_{e} v_{Te})^{2}$, $g=u/v_{Te}$, $u$ is the relative velocity of the scattering particles, and $v_{Te}$ is thermal velocity of electrons. 

In Eq. (\ref{Xi}), the function $G(g)$ takes into account the Fermi-Dirac statistics of electrons  and represents the relative availability of states that contribute to the scattering \cite{Rightley}:

\begin{equation} \label {functionG}
G(g)= {\frac{\eta \exp^{-g^{2}}g^{5}} {[-Li_{3/2}(-\eta)](\eta e^{-g^{2}}+1)^{2}}}.
\end{equation}
where $-{\rm Li}_{3/2}=[-\eta]=\frac{4}{3\sqrt{\pi}} \theta^{-3/2}$, $\eta=\exp(\mu/k_{B}T_e)$, and $\mu$ is the electron chemical potential \cite{Melrose, Rightley}.

The  transport cross-section $Q^{T}$ is computed according to two-particle quantum scattering theory \cite{Massey1965}:
\begin{equation}\label{Transport}
Q^{T} (k)=\frac{4\pi }{k^{2} } \sum _{l}(l+1)\sin ^{2}  (\delta _{l} (k)-\delta _{l+1} (k)),
\end{equation} 
the phase shift  $\delta _{l} (k)\equiv \delta _{l} (k,r\to \infty )$ is calculated by solving the Calogero equation ~\cite{Calogero, Babikov, Babikov0}:
\begin{equation}\label{Phase} 
    \frac{d\delta _{l} (kr)}{dr} =-\frac{1}{k} \Phi (r)\left[\cos \delta _{l} (kr)j_{l} (k,r)\right.-\left.\sin \delta _{l} (k,r)n_{l} (k,r)\right]^{2},
\end{equation} 
with the condition $\delta _{l} (k,0)=0$.
In Eq.~(\ref{Phase}), $k=mu/\hbar$ is the wave number, $l$ indicates the orbital quantum number, $j_{l}$ and $n_{l} $ are the Rikkati-Bessel functions, and $\Phi(Y)$ is the pair interaction potential between colliding particles. In addition, for the electrical conductivity we compare the NPS model results with other theoretical approaches as well as with the data from the density functional theory simulations. 

\subsection{Temperature relaxation and conductivity}

The relaxation rates of the electron and ion temperatures are determined by the  collision rates (frequencies) between electrons and ions and the difference of  electron and ion temperatures  ~\cite{Glosli}:
\begin{equation} \label{dT}
\frac{dT_{e} }{dt} =\frac{T_{i} -T_{e}}{\tau_{ei}},\,\,
\frac{dT_{i} }{dt} =\frac{T_{e} -T_{i}}{\tau_{ie}},
\end{equation}
where the collision frequency reads:
\begin{equation} \label{nu}
\nu_{ei} = \frac{1}{\tau_{ei}} = \frac{8\sqrt{2\pi} n_{i}Z^{2}e^{4} \Xi}{3 m_{e} m_{i}} \left(\frac{k_{B}T_{e}}{m_{e}}+\frac{k_{B}T_{i}}{m_{i}}\right)^{-3/2}\, 
\end{equation}
\begin{equation} \label{nuie}
\nu_{ie} = \frac{1}{\tau_{ie}}= Z \nu_{ei}.
\end{equation}

The electrical conductivity is another important transport coefficient that we consider in this work.  We consider a dense plasma regime where the resistivity is dominated by the scattering of electrons on ions.  Electrical conductivity is defined in terms of the electron-ion collision frequency as \cite{Rightley}
\begin{equation} \label {Omega}
 \Omega= {\frac{e^2 n_{e}}{m_{e} \nu_{ei}}}.
\end{equation}

We compute the generalized Coulomb logarithm, the temperature relaxation rates, and electrical conductivity using NPS model. In addition, we compare the NPS model results with the ideal plasma model with $\gamma=0$ and with the case where ionic non-ideality is neglected by setting $k_i=\lambda_i^{-1}$. 

\section{\label{Sec. III} Results and discussions}

In Fig. \ref{Fei} shows the effective potential of electron-ion interaction  Eq. (\ref{Yukawa}) with the NPS model for screening. We compare the NPS results with the calculations performed setting $\gamma=0$. In Fig. \ref{Fei}, the results for (a) $r_{s}=0.5$, (b) $r_{s}=1$, and (c) $r_{s}=3$ are presented at the degeneracy parameter values  $\theta=0.1$ and $\theta=1$. In Fig. \ref{Fei}, the solid line indicates the results calculated using  $\gamma=0$, and the dotted line indicates the results when $\gamma\neq0$. As can be seen in the Fig. \ref{Fei}, the correction to electron screening due to electron exchange correlations noticeably affects the effective potential at small degeneracy parameters. Here, we note that the importance of the difference in the considered screened potentials depends on the energy of the collision during electron-ion scattering.
Therefore, from Fig. \ref{Fei}, one cannot gauge to what degree the electronic exchange-correlation effects are essential for plasma transport properties. From Fig. \ref{Fei}, we can conclude that the exchange-correlation effects lead to a stronger screening of the electron-ion interaction \cite{Moldabekov2}.

In Fig.\ref{gamma}, the dimensionless coefficient $\gamma$ entering the NPS model is shown as a function of the degeneracy parameter $\theta$ at different densities.
We can conclude from  Fig.\ref{gamma} that with the increasing density and degeneracy parameters, the effect on the screening properties of plasma electrons becomes weaker.

Fig. \ref{ksfigs} shows the inverse screening lengths $k_e$ (panel (a)) and $k_s$ (panel (b)) in atomic units as the function of the degeneracy parameter $\theta$ with and without taking into account exchange-correlation effects for different parameters $r_{s}$. 
From Fig. \ref{ksfigs}, one can see that the impact of the electron exchange correlations on the screening length is relevant for $r_s\gtrsim 1$. 

Next, we analyze the effect of the correction to electron screening due to electron exchange correlations by considering the transport cross-section for electron-ion scattering. The corresponding results are presented in Fig. \ref{crosssections}, where we show the transport cross sections for electron scattering on a proton at the degeneracy parameter $\theta=1.0$. Calculations were carried out by solving the Calogero equation Eq.~(\ref{Phase}). It is shown that taking into account electronic nonideality leads to an increase in the scattering cross sections at low densities with $r_s\gtrsim 1$ and low collision velocities with $v<v_{\rm th}$. In addition, we see that the effect of electronic non-ideality on the transport scattering cross section reduces with the increase in the degeneracy parameter. From Fig. \ref{crosssections}, we see that at considered densities, the ideal electron approximation for screening becomes accurate at $\theta=3$.

\begin{figure}[t]
\includegraphics[width=0.9 \linewidth]{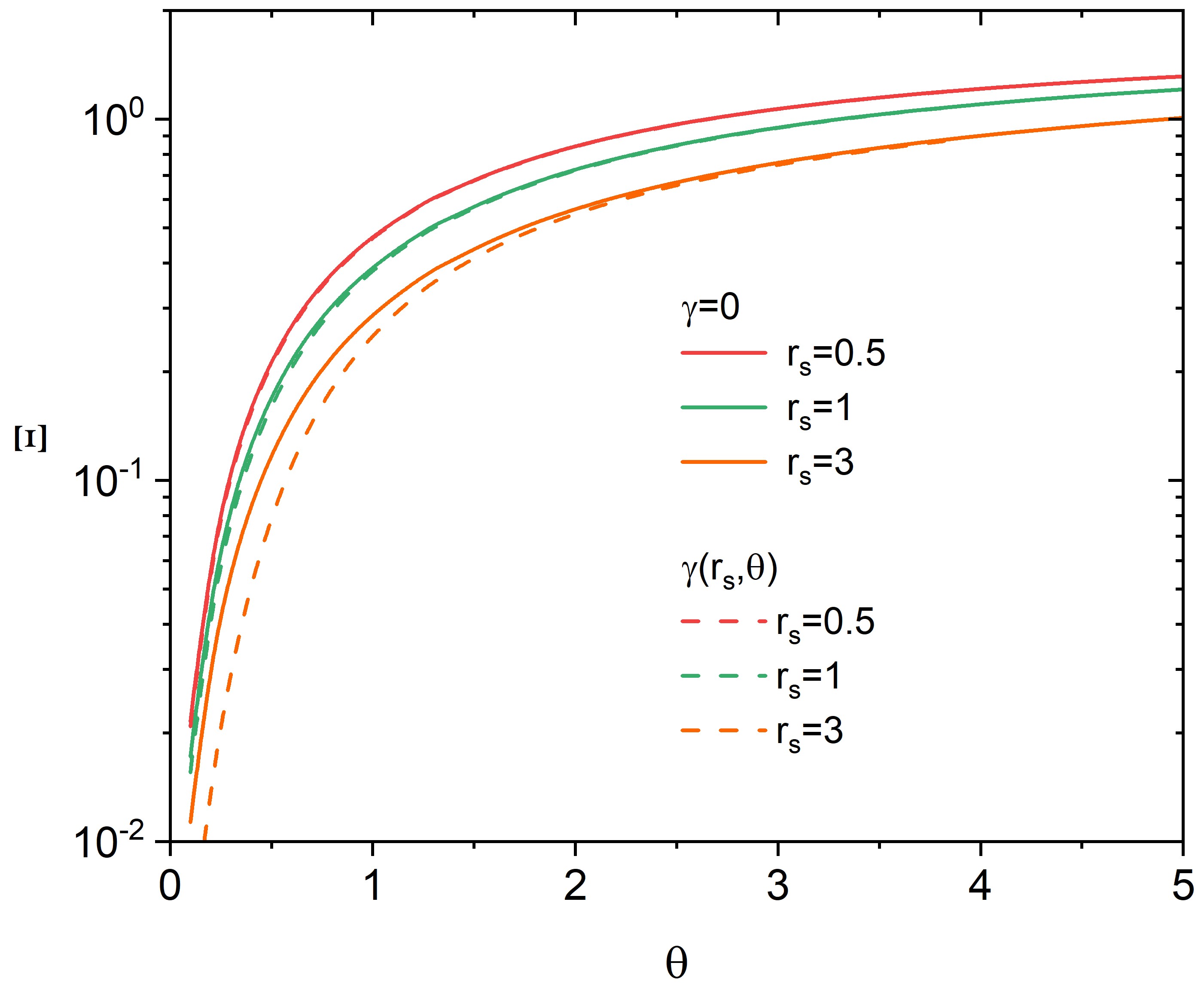}
\caption{The Coulomb logarithm at the degeneracy parameter $\theta$ for different values of the density parameter $r_{s}$ for cases with and without the electronic non-ideality correction $\gamma$.}
\label{Coulomblog}
\end{figure}

Fig. \ref{Coulomblog} shows the generalized Coulomb logarithm  (\ref{Xi}) as a function of $\theta$. From Fig. \ref{Coulomblog}, one can observe that the electronic exchange correlations can be safely neglected at $r_s=0.5$.
The effect of electronic nonideality starts to become noticeable at $r_s\geq 1$. 
From Fig. \ref{Coulomblog}, we see that taking into account the electronic local field correction (i.e., $\gamma$) leads to smaller values of the Coulomb logarithm in the transition region from nondegenerate to degenerate plasma at $\theta<3$.

\begin{figure}[t]
\includegraphics[width=0.9 \linewidth]{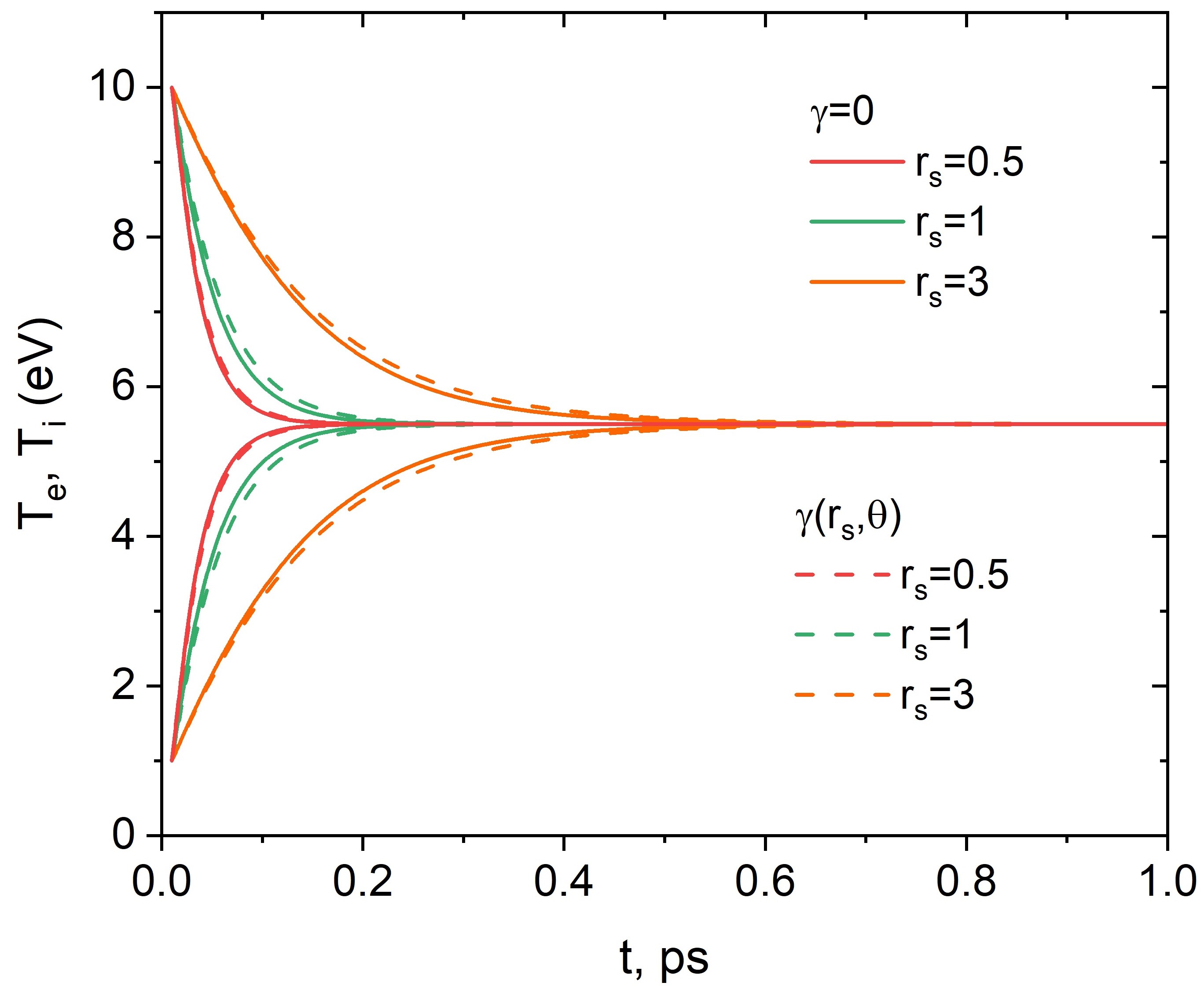}
\caption{The temperature relaxation time between electrons and ions with $\gamma\neq0$ and without $\gamma=0$ correction for different densities.}
\label{TeTi}
\end{figure}

Now, we investigate the effect of electronic local field correction on temperature relaxation in a hydrogen plasma with electrons hotter than the ions. For that we solve Eqs.~(\ref{dT}-\ref{nu}).
Based on the calculation of the Coulomb logarithm $\Xi$, the temperature relaxation times were computed for different densities with the initial temperatures of electrons $T_e=10~{\rm eV}$ and of ions $T_e=1~{\rm eV}$.
The results for the temperature relaxation are presented in Fig. \ref{TeTi}, where we compare the NPS model results with the data computed by setting  $\gamma=0$. With increasing density, the frequency of collisions between the electrons and ions increases, leading to faster equalization of the temperatures of electrons and ions. From Fig. \ref{TeTi}, it is clear that the lower the density, the more time it takes for the system to reach thermodynamic equilibrium.
The electron change correlations effect leads to longer relaxation times.
This can be explained by recalling that the electronic nonideality leads to a stronger screening of the electron-ion pair potential, which reduces the collision rates between electrons and ions.

\begin{figure}[h]
\includegraphics[width=0.9 \linewidth]{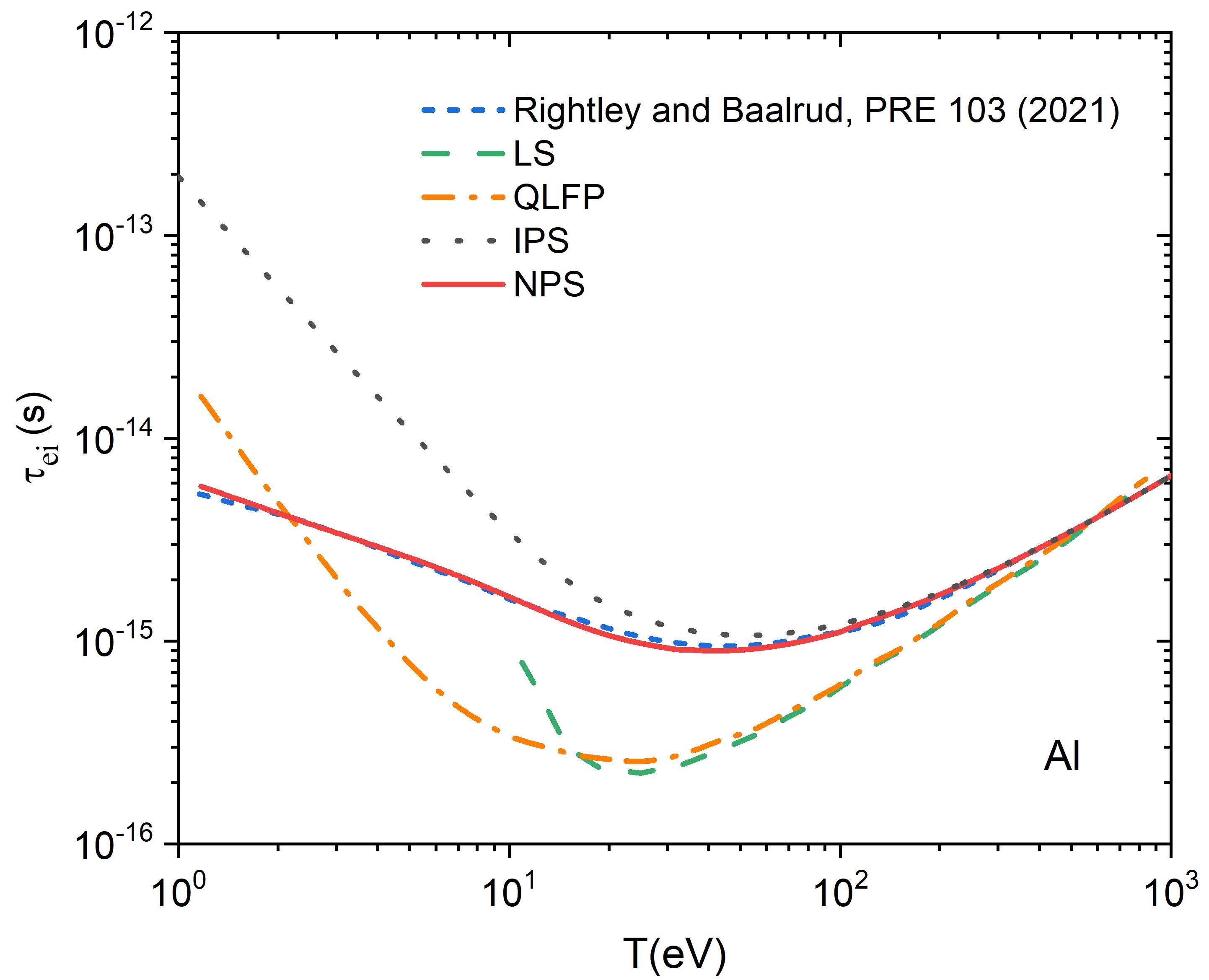}
\caption{Electron-ion collisional relaxation rate as a function of temperature in solid density ($\rho=2.7 g cm^{-3}$) aluminum: solid red line NPS,  dotted black line IPS models; short dashed blue line Rightley, Baalrud model \cite{Rightley}; dashed green line LS model \cite{Spitzer56}; dash-dotted orange line QFLP model \cite{Daligault16,Daligault_QLFP}.}
\label{Relax}
\end{figure}

As another example of applying the NPS model, in Fig. \ref{Relax}, we show the relaxation rate (i.e., inverse collision frequency) computed for warm dense aluminum at solid density. We compare the NPS model results with the data obtained using the IPS model, with the results of the well-known Landau-Spitzer model \cite{Spitzer56}, and with the results of advanced state-of-the-art methods, including quantum Landau-Fokker-Planck (QLFP) model results from  \cite{Daligault16, Daligault_QLFP} and the average-atom model based model developed recently by Rightley and Baalrud \cite{Rightley}.

As one might expect, from Fig. \ref{Relax}, we see that all models provide similar values at high temperatures due to the weakening of the correlation effects. In contrast, at low temperatures, the models that include strong correlation effects and the models designed for weakly coupled plasmas (or ideal plasmas) start to differ from each other with the decrease in temperature.
In non-degenerate plasma with $T>100~{\rm eV}$, the relaxation time can be calculated using the ideal plasma model for screening. At a high and partial degeneracy with $T\lesssim 10~{\rm eV}$, both quantum effects and plasma nonideality must be taken into account. 
Therefore, as shown in Fig. \ref{Relax}, the consideration of the exchange-correlation effects is crucial for the WDM regime. 
We found that the NPS model agrees with the results of  Rightley and Baalrud \cite{Rightley}.

\begin{figure}[h]
\includegraphics[width=0.9 \linewidth]{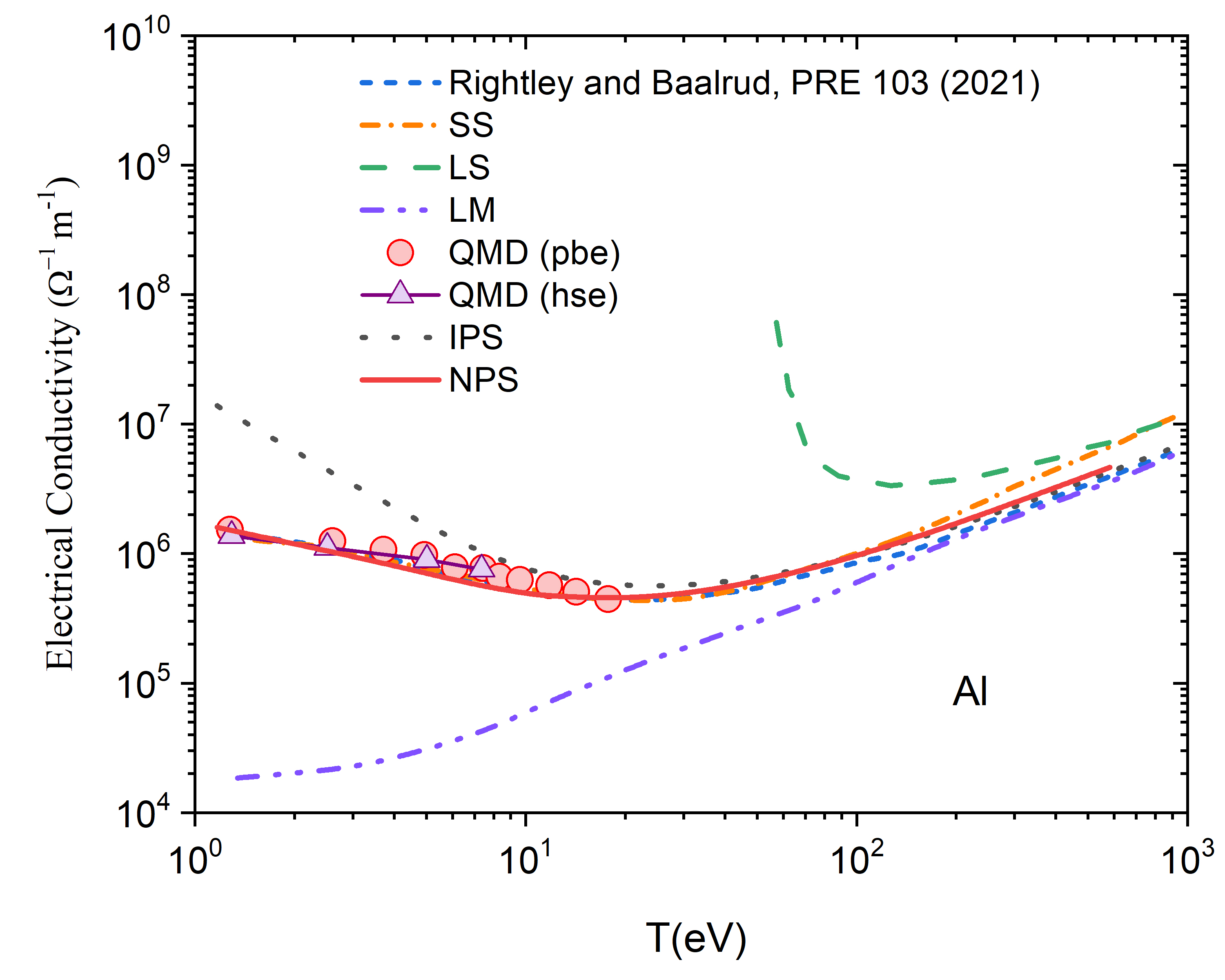}
\caption{{Electrical conductivity as a function of temperature in solid density ($\rho=2.7 g cm^{-3}$) aluminum at different temperatures.  Solid red line corresponds to the NPS model, dotted black line indicates the IPS model results, the Starrett and Shaffer (SS) results are shown  by dot-dashed orange line \cite{Shaffer20}, the Lee-More model results are depicted by dot-dashed purple line \cite{Lee}, the LS model data corresponds to dashed green line, and  the QMD results of Witte et al. \cite{Witte} using the Perdew–Burke-Ernzerhof and Heyd–Scuseria–Ernzerhof exchange-correlation functionals are shown by symbols.}
}
\label{Electricalconductivity}
\end{figure}

As a third example of the application of the NSP model, in  Fig. \ref{Electricalconductivity}, we show the electrical conductivity of aluminum at density $\rho=2.7 g cm^{-3}$ and different temperatures. In addition to the aforementioned models, here we compare with the results from the model of Starrett and Shaffer (SS) \cite{Shaffer20}, the Lee-More model (LM) \cite{Lee},  and the quantum molecular dynamics (QMD) results of Witte \textit{et al.} \cite{Witte} obtained using the exchange-correlation functionals of Perdew–Burke–Ernzerhof (pbe) and Heyd–Scuceria–Ernzerhof (hse). From Fig. \ref{Electricalconductivity}, one can see that the NPS model is in good agreement with more elaborated calculations results from QMD \cite{Witte},  from Ref. \cite{Shaffer20} (SS), and the results by Rightley and Baalrud \cite{Rightley}. The IPS model overestimates conductivity at $T\lesssim 10~{\rm eV}$. 


\section{Conclusion}

Using screening theory in quantum dense plasmas, we proposed a simple model for calculating the relaxation and transport properties of dense plasmas. 
The proposed model uses quantum Monte Carlo results for the exchange-correlation free energy density and the ionic non-ideality in the screening length of plasmas. This screening length is used in Yukawa-type pair interaction potential of particles and combined with the theory of generalized Coulomb logarithm. The comparison of the results with the data from state-of-the-art methods shows that the proposed model can be used to evaluate the transport and relaxation properties of warm, dense matter. We note that the presented model is limited to systems with a high degree of ionization. 

\section*{Acknowledgments}
This research is funded by the Science Committee of the Ministry of Education and Science of the Republic of Kazakhstan Grant AP19678033 "The study of the transport and optical properties of hydrogen at high pressure".

\bibliography{Main}
 
\end{document}